\documentclass[aps,prl,showpacs,twocolumn,preprintnumbers]{revtex4}

\usepackage{amssymb,amsmath}
\usepackage{graphicx}
\usepackage{epsfig}
\usepackage{dcolumn}
\usepackage{bm}
\usepackage{citesort}
\usepackage{color}

\newcommand{\braket}[2]{\langle #1 | #2 \rangle }

\def\rvec{ \vec{r} }
\def\kvec{ \vec{k} }

\def\simlt{\mathrel{\lower .3ex \rlap{$\sim$}\raise .5ex \hbox{$<$}}}
\def\simgt{\mathrel{\lower .3ex \rlap{$\sim$}\raise .5ex \hbox{$>$}}}

\begin{document}

\title{Exchange in a silicon-based quantum dot quantum computer architecture}
\author{S.N. Coppersmith}
\affiliation{Department of Physics, University of Wisconsin, Madison, Wisconsin 53706}
\author{Seungwon Lee}
\author{Paul von Allmen}
\affiliation{Jet Propulsion Laboratory, California Institute of Technology,
Pasadena, California  91109}

\begin{abstract}
In bulk silicon,
intervalley electronic interference
has been shown to lead to strong oscillations in the exchange coupling between
impurity electronic wavefunctions, posing
a serious manufacturability problem for proposed quantum computers.
Here we show that this problem does not arise in
proposed architectures using Si/SiGe quantum dots because of the
large in-plane strain in Si quantum wells together with the
strong confinement potential typical of heterostructures.
\end{abstract}

\pacs{03.67.Lx, 73.21.Fg, 73.21.La}
\maketitle
{\bf Introduction.}  The proposal to build
a quantum computer with qubits that are single electrons
confined in gated silicon quantum dots in silicon/silicon-germanium
heterostructures~\cite{vrijen00,friesen03} is attractive because
of the self-aligning properties of quantum dots as well as the long
coherence times of electron spins
in silicon~\cite{feher59,tyrishkin03}.
However, one complication of silicon compared to a direct band gap
material such as gallium arsenide is the
valley degeneracy in the band structure.
As Refs.~\cite{koiller02a,koiller02b,wellard03,koiller03} point out,
interference between valleys
causes charge oscillations that can cause the coupling between qubits to
vary rapidly in space.
Since the fundamental two-qubit gate is achieved by varying the overlap of
electron wavefunctions centered at different locations, obtaining reliable gates
would appear to require controlling the individual wavefunctions on the
scale of the oscillations and not the much longer length scale describing the
variation of the envelope of the wavefunctions.
This sensitivity has severe implications for the feasibility of
constructing a reliable quantum computer using silicon.

Here we present theoretical arguments and atomistic simulations
that show that
the large in-plane strain present in a silicon quantum well, together
with the strong potential confining electrons in the quantum well,
combine to eliminate the problem
of the fast oscillations in the overlap of wavefunctions of electrons
centered at different locations in a Si quantum well
in a Si/SiGe heterostructure.
The result depends on the strong heterostructure confinement present
in the gated quantum dot architecture, and does not apply to other
Si-based quantum computing schemes~\cite{kane98,obrien01,ladd02}.

{\bf Effect of Strain on Electron Wavefunctions in Si Quantum Wells.}
Unlike gallium arsenide,
unstrained silicon has six-fold degenerate conduction band minima located
along the [100], [010], and [001] directions, about 85\% of the way
to the boundary of the Brillouin zone.
Interference between these valleys causes fast oscillations
in the electron density,
on the scale of the interatomic spacing.
In bulk silicon the oscillations occur along the $x$, $y$, and $z$ directions.
These fast oscillations make controlling the exchange
coupling accurately, which is necessary
to make a two-qubit gate~\cite{loss98},
much more difficult~\cite{koiller02a,koiller02b,koiller03}.

In Si/SiGe heterostructures the
large in-plane strain present in the quantum well
reduces the six-fold
valley degeneracy to a four-fold and a two-fold one. For a typical quantum 
well in the x-y plane,
perpendicular to the [001] crystal axes,
 the two lowest
energy valleys are at $k_x=0$, $k_y=0$, and $\pm k_z$, with $k_z \ne 0$.
For typical heterostructures, the
minima with nonzero $k_x$ or $k_y$ have energies more than 0.1 eV higher
than the minima with nonzero $k_z$, so that at the low temperatures
at which a quantum computer would operate, only the valleys with
$k_x = k_y = 0$ are relevant.~\cite{tahan02,koiller02b,koiller03}
Therefore, in the plane of the quantum well the charge density does
not oscillate.
However, the two degenerate valleys along $\pm k_z$ do lead to oscillations
along the $z$ direction.
Therefore, strain alone does not completely remove the extreme
sensitivity of the exchange coupling to small changes
in position, especially for impurity based qubits~\cite{koiller02b,koiller03}.

Below we show that
in a Si/SiGe quantum dot quantum computer
wavefunction oscillations along $z$ do not
lead to problems with controllability of the exchange interaction because
the oscillations of the wavefunctions of different qubits are aligned by
the strong confinement potential of the quantum well.
The argument is quite general, and the result is robust even in the presence
of gate potential fluctuations and imperfections such
as quantum well width variations.

{\bf Role of Confinement Potential.}
The exchange interaction does not exhibit fast
oscillations in quantum wells because the oscillations along $z$
are aligned by the strong quantum well potential.
The physical reason for the alignment of the charge density oscillations
perpendicular to the quantum well plane is that the quantum
well confinement potential varies on a much shorter length scale
than any in-plane potential variations.
(Recall that typical quantum wells have
10 nm widths and potential steps
occurring on the single unit cell scale and heights
$\simgt$ 0.1 eV~\cite{schaffler97}, 
while the potentials defining the quantum dot vary on length scales of many tens of nanometers.)
Because of this
separation of length scales, it is appropriate to use
the Born-Oppenheimer (B-O) approximation~\cite{born27,bentosela98}.
We wish to find the lowest energy eigenstate $\psi_{E}$ of
the time-independent
Schr\"odinger equation
of an electron in a single quantum well,
\begin{eqnarray}
\left [- \frac{\hbar^2}{2m} \left ( \frac{\partial^2}{\partial x^2}
+ \frac{\partial^2}{\partial y^2}+\frac{\partial^2}{\partial z^2} \right )
+ V(x,y,z) \right ]\psi_E(x,y,z) \nonumber \\
= E\psi_E(x,y,z)~.
\label{eq:schrodinger_envelope}
\end{eqnarray}
We write the potential $V(x,y,z)$ as the sum of three contributions,
the atomic potential $V_a(x,y,z)$, the confinement potential $V_c(x,y,z)$,
and the residual potential $V_r(x,y,z)$, which accounts for
potentials from external gates and possible
weak impurity potentials (e.g., from dopant inhomogeneities in the
modulation doping layer).
For simplicity, initially we will ignore the atomic potential $V_a(x,y,z)$
and consider a system with confinement and residual potentials whose
sum varies much quickly along the $z$ direction than
in the $x-y$ plane.
We follow
the usual B-O argument~\cite{born27,chandler87}
and write $\psi(x,y,z) = \Phi_0(z; x,y)\chi(x,y),$ where
$\Phi_0(z; x,y)$ describes the ground state wavefunction as a
function of $z$ at given $x$ and $y$, and $\chi(x,y)$ describes
the in-plane variations.
The B-O prescription is to first find
$\Phi_0$ by solving
\begin{equation}
\left [- \frac{\hbar^2}{2m} \frac{d^2}{dz^2} + V(x,y,z) \right ]
\Phi_0(z; x,y)
= V_{eff}(x,y)\Phi_0(z; x,y),
\label{eq:z_dependence}
\end{equation}
and then determine $\chi(x,y)$ from
\begin{equation}
\left [- \frac{\hbar^2}{2m} \left ( \frac{\partial^2}{\partial x^2}
+ \frac{\partial^2}{\partial y^2}\right )
+ V_{eff}(x,y) \right ]
\chi(x,y)
= E\chi(x,y).
\label{eq:xy_dependence}
\end{equation}

Eq.~(\ref{eq:z_dependence}) for the $z$-dependence of the
wavefunction depends only on the local value of $x$ and $y$.
The wavefunctions obtained for two different residual potentials
in the $x-y$ plane, $V_{r1}(\rvec)$ and $V_{r2}(\rvec)$
(corresponding to electrons centered on two different quantum dots),
are $\Phi_0(z; x,y)\chi_1(x,y)$ and $\Phi_0(z; x,y)\chi_2(x,y)$.
Here, $\chi_1$ and $\chi_2$ are distinct, but $\Phi_0$ is the same
for the two wavefunctions.
Therefore, while the wavefunction can vary quickly as a function of $z$,
this $z$-dependence is identical for wavefunctions
centered at different locations in the $x-y$ plane.
The overlap matrix element between electrons centered at
$\vec{r}_1$ and $\vec{r}_2$,
$\braket{{\vec{r}_1}}{{\vec{r}_2}}$
is
\begin{eqnarray}
\braket{\rvec_1}{\rvec_2}
= \int dx dy\ \chi_1^*(x,y)\chi_2(x,y)
\left [ \int dz |\Phi_0(z; x,y)|^2 \right ].
\label{eq:overlap}
\end{eqnarray}
The term in brackets does not depend on the residual
potential $V_r(x,y)$, so it does not change when the gate potentials
are varied.
In other words, the strong quantum well confinement ``locks" the
variations perpendicular to the quantum well plane.
Since, as discussed above, the wavefunctions in these strained
quantum wells have no oscillations
in the $x-y$ plane, the scale of all variations in the exchange
coupling is the quantum
dot size, typically of order at least 100 nm.
Therefore, the valley degeneracy does not lead to additional
problems in controlling the exchange interaction in a Si/SiGe quantum dot
computer.

To include the atomic potential $V_a(x,y,z)$, we use
the envelope approximation~\cite{kohn57} and
write the wavefunction 
\begin{equation}
\psi(\rvec) = \sum_{j=1}^2 \alpha_j F_j(\rvec)u_j(\rvec)e^{i\kvec_j\cdot\rvec_j}~,
\label{eq:envelope_approx}
\end{equation}
where $u_j(\rvec)e^{i\kvec_j\cdot\rvec_j}$ is the Bloch wave at the
minimum of the $j^{th}$ valley, the $\alpha_j$ describe the amplitude
of the contribution from each valley, and the $F_j(\rvec)$ are ``envelope"
functions.
Because band offsets in typical heterostructures
are $\simlt 0.2$~eV, while atomic potentials are several eV,
the variations induced by the
confinement potential are much slower than those induced by
the atomic potential, and therefore the envelope function
itself satisfies the Schr\"odinger equation,
Eq.~(\ref{eq:schrodinger_envelope})~\cite{kohn57}.
Thus, within the envelope approximation,
the argument given above with no atomic potential
applies with no modifications.

In conclusion, we have shown that the strong confinement potential in Si/SiGe
quantum wells causes the exchange coupling to depend smoothly
on quantum dot separation.

{\bf Tight-Binding Model Calculation.}
In explicit support of the aforementioned conclusions we have numerically computed
the behavior of the exchange coupling
as a function of the separation between quantum dots in Si/SiGe
heterostructures by using the quantitative nano-electronic modeling tool
NEMO-3D~\cite{nemorefs}.  NEMO-3D describes the electron Hamiltonian 
in the framework of an $sp^3d^5s^*$ nearest-neighbor empirical tight-binding model,
which allows us to
incorporate explicitly the effect of well-width variation, 
interface roughness, and strain at the atomic level.
The atomistic description is essential to verify the aforementioned argument
that the electron wave function varies smoothly in the $x-y $ plane 
without fast oscillations at the atomic scale.
We use
the empirical tight-binding
parameters of Ref.~\cite{boykin_tb}, which reproduce
both the band edges and effective masses of the lowest conduction
band and the highest valence band to within less than 5\%.
The effects of strain on the band edges and effective masses
are incorporated into the model by modifying the tight-binding parameters 
with the L\"owdin orthogonalization procedure,
the Slater-Koster table, and the generalized version of 
the Harrison's $d^{-2}$ scaling law~\cite{boykin_strain}.  

\begin{figure}[t]
\scalebox{0.4}{\includegraphics*{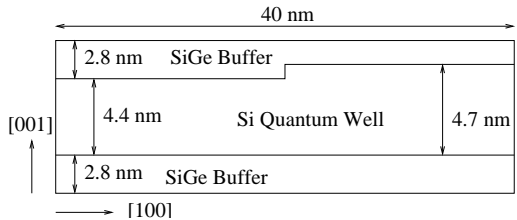}}
\caption{Schematic cross section of the modeled system
which consists of a Si quantum well with a width of 4.4~nm
and a Si$_{0.7}$Ge$_{0.3}$ barrier material 
with a width of 2.8~nm below and above the Si well.
The width of one half of the Si well is increased
by one monolayer to imitate a well-width variation
arising from growth inhomogeneity.} 
\label{fig:simulation}
\end{figure}

Figure~\ref{fig:simulation} shows the modeled system,
a strained Si quantum well with a width of 4.4~nm and a relaxed Si$_{0.7}$Ge$_{0.3}$ 
barrier buffer with a width of 2.8~nm below and above the Si well.
The supercell dimension in the $x$-$y$ plane is 40~nm. 
To simulate the effect of the well-width variation due
for example
to a miscut substrate,
the Si well width is
increased by one monolayer in half the
supercell in the x direction.
The
external gate potential $V_g(x,y,z)$ for the lateral
confinement is approximated by
a harmonic potential in the $x-y$ plane and a linear potential along the $z$ direction: 
$V_g(x,y,z)=A (x^2+y^2)+ B z$,
with $A$ and $B$ chosen to be 0.1~meV/nm$^2$ and 20~meV/nm,
respectively, in agreement with
electrostatic potential calculations \cite{friesen}.
All other potentials $V_c(x,y,z)$ and $V_a(x,y,z)$
are included in the tight-binding Hamiltonian
via the couplings between the tight-binding basis orbitals.
By diagonalizing the resulting Hamiltonian,
the ground state electron wave function $\psi(\vec{r})$
for a single quantum dot is obtained.
The two-electron wave functions in a system
of two quantum dots
are prepared by superposing the single electron wave functions centered
at each of the two quantum dots. 
Two types of two-electron wave functions can be constructed:
symmetric and antisymmetric states. 

The exchange coupling $J$ of the two-electron system or, 
equivalently,
the energy difference between the symmetric and antisymmetric states
is given approximately by~\cite{koiller02a}
\begin{eqnarray}
&~&J(\vec{R}) =
\int dx_1\ dy_1\  dz_1\ dx_2\  dy_2\  dz_2 
\nonumber \\
&~&\psi^*(\vec{ r_1}) \psi^*(\vec{r_2}-\vec{R}) 
\frac{e^2}{\epsilon|\vec{r_1}-\vec{r_2}|}\  \psi(\vec{r_1}-\vec{R})\ \psi(\vec{r_2}), 
\end{eqnarray}
where $\vec{R}$ is the relative distance vector
between the centers of the two quantum dots
and $\epsilon$ is the dielectric constant.~\cite{good_approx}
This exchange integral is further expanded
into integrals involving tight-binding basis orbitals. 
The details of the expansion and the calculation of the
integrals can be found elsewhere~\cite{lee_exchange}. 

\begin{figure}[t]
\scalebox{0.55}{\includegraphics*{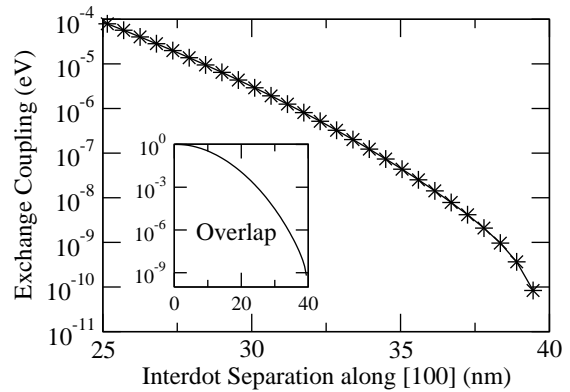}}
\caption{Exchange coupling as a function of the interdot separation
along the [100] direction for two quantum dots that are 
electrostatically defined in a Si quantum well surrounded by a
Si$_{0.7}$Ge$_{0.3}$ buffer. The inset shows the overlap between
single-electron wave functions $\psi({\bf r})$ centered at each
of the two quantum dots, i.e.
$|\int dx\ dy\ dz\ \psi^*(\vec{r}-X) \psi(\vec{r})|^2$,
as a function of the interdot separation $X$ along the [100] direction.
Both the exchange coupling and the overlap
(inset) smoothly increase
as the interdot separation decreases. This result clearly demonstrates
the absence of atomic-level oscillations.
} 
\label{fig:exchange}
\end{figure}

Figure~\ref{fig:exchange} shows the exchange coupling as a function
of the interdot separation along the [100] direction. 
As the interdot separation decreases, the exchange coupling smoothly
increases without atomic-level oscillations.
To show the origin of this behavior, we also plot the overlap between
single-electron wave functions centered at each of the two quantum dots
in the inset of Fig.~\ref{fig:exchange}.
As expected, the overlap smoothly increases with the decrease of the
interdot separation.
These results clearly demonstrate the absence of atomic-level
oscillations in the single-electron wave function, and consequently
the absence of fast oscillations in the exchange coupling
in the two-electron wave function.

{\bf Summary.}  We have shown that the exchange interaction between two
qubits composed of quantum dots in a Si/SiGe heterostructure exhibits a smooth dependence
on qubit separation.  The origin of the difference between the behavior in quantum dots
and in bulk silicon impurities is the strain and the strong confining potential in the heterostructure.

{\bf Acknowledgements.}
We thank the entire UW Solid State Quantum Computing group, and
particularly Mark Eriksson, Mark Friesen, Bob Joynt, and Don Savage, for extremely useful
conversations. We also thank all the developers of NEMO-3D for their contributions 
to this quantum modeling tool. We gratefully acknowledge support from the NSF under
Grants DMR-0209630 and DMR-0325634, and by ARDA, ARO, and NSA.


\begin{thebibliography}{99}
\bibitem{good_approx}
  The validity of this approximation has been verified in detail via
  ab initio simulations in Ref.~\cite{koiller03}.
\end{thebibliography}
\end{document}